\documentclass[english,useAMS,usenatbib]{mn2e}

\usepackage{babel}
\usepackage{mathptmx}
\usepackage[T1]{fontenc}
\usepackage[latin9]{inputenc}
\usepackage[]{natbib}
\usepackage[]{url}

\usepackage{graphicx}
\usepackage{txfonts}

\newcommand{\refa}{$^\mathrm{(a)}$}
\newcommand{\refb}{$^\mathrm{(b)}$}

\newcommand{\refc}{$^\mathrm{(c)}$}
\newcommand{\refd}{$^\mathrm{(d)}$}
\newcommand{\refe}{$^\mathrm{(e)}$}

\newcommand{\refdf}{$^\mathrm{(d,f)}$}
\newcommand{\reff}{$^\mathrm{(f)}$}
\newcommand{\refg}{$^\mathrm{(g)}$}
\newcommand{\refh}{$^\mathrm{(h)}$}
\newcommand{\refi}{$^\mathrm{(i)}$}
\newcommand{\refj}{$^\mathrm{(j)}$}

\newcommand{\MBa}{$\mathrm{M}_{\mathrm{Ba}}$~}
\newcommand{\MBb}{$\mathrm{M}_{\mathrm{Bb}}$~}
\newcommand{\MBaa}{$\mathrm{M}_{\mathrm{Baa}}$~}
\newcommand{\MBab}{$\mathrm{M}_{\mathrm{Bab}}$~}

\newcommand{\MTot}{$\mathrm{M}_{\mathrm{Tot}}$~}

\newcommand{\MJ}{$\mathrm{M}_{\mathrm{J}}$~}
\newcommand{\MS}{$\mathrm{M}_{\odot}$~}

 \title[Lucky Imaging Adaptive Optics of GJ569Bab]{Lucky Imaging Adaptive Optics of the brown
   dwarf binary GJ569Bab\thanks{Based on service observations made with the WHT telescope operated on the island of La
   Palma by the Isaac Newton Group and on observations made with the Nordic Optical Telescope, operated on the island of
   La Palma jointly by Denmark, Finland, Iceland, Norway, and Sweden, in the Spanish Observatorio del Roque de los
   Muchachos of the Instituto de Astrof\'{\i}sica de Canarias.}}

\author[B.~Femen\'{\i}a et al.]{B.~Femen\'{\i}a$^{1,2}$\thanks{E-mail:bfemenia@iac.es}, R.~Rebolo$^{1,3}$,
  J.~A.~P{\'e}rez-Prieto$^{1}$, S.~R.~Hildebrandt$^{1,4}$, L.~Labadie$^{1,2}$, \newauthor A.~P{\'e}rez-Garrido$^{5}$,
  V.~J.~S.~B{\'e}jar$^{1,2}$, A.~D\'{\i}az-S{\'a}nchez$^{5}$, I.~Vill{\'o}$^{5}$, A.~Oscoz$^{1}$, R.~L{\'o}pez$^{1}$,
  \newauthor L.~F.~Rodr\'{\i}guez$^{1}$ and J.~Piqueras$^{1,6}$ \\ $^{1}$Instituto de Astrof\'{\i}sica de Canarias, C/
  V\'{\i}a L{\'a}ctea S/N, E-38200 La Laguna, Spain\\ $^{2}$Departamento de Astrof\'{\i}sica, Universidad de La Laguna,
  E-38205 La Laguna, Tenerife, Spain\\ $^{3}$Consejo Superior de Investigaciones Cient\'{\i}ficas,
  Spain\\ $^{4}$Laboratoire de Physique Subatomique et Cosmologie, F-38000 Grenoble, France\\ $^{5}$Universidad
  Polit{\'e}cnica de Cartagena, Campus Muralla del Mar, Cartagena, Murcia E-30202, Spain\\ $^{6}$ Max-Planck-Institut
  f\"ur Sonnensystemforschung, Max-Planck-Str. 2, D-37191 Katlenburg-Lindau, Germany}
\begin{document}

\date{Accepted 2010 December 16. Received 2010 December 15; in original form 2010 August 2}

\pagerange{\pageref{firstpage}--\pageref{lastpage}} \pubyear{2010}

\maketitle

\label{firstpage}

\begin{abstract}
The potential of combining Adaptive Optics (AO) and Lucky Imaging (LI) to achieve high precision astrometry and
differential photometry in the optical is investigated by conducting observations of the close 0\farcs1 brown dwarf
binary GJ569Bab. We took 50000 $I$-band images with our LI instrument FastCam attached to NAOMI, the 4.2-m William
Herschel Telescope (WHT) AO facility. In order to extract the most of the astrometry and photometry of the GJ569Bab
system we have resorted to a PSF fitting technique using the primary star GJ569A as a suitable PSF reference which
exhibits an $I$-band magnitude of $7.78\pm0.03$. The AO+LI observations at WHT were able to resolve the binary system
GJ569Bab located at $4\farcs 92 \pm 0\farcs05$ from GJ569A. We measure a separation of $98.4 \pm 1.1$ mas and $I$-band
magnitudes of $13.86 \pm 0.03$ and $14.48 \pm 0.03$ and $I-J$ colors of 2.72$\pm$0.08 and 2.83$\pm$0.08 for the Ba and
Bb components, respectively.  Our study rules out the presence of any other companion to GJ569A down to magnitude
I$\sim$ 17 at distances larger than 1\arcsec. The $I-J$ colors measured are consistent with M8.5-M9 spectral types for
the Ba and Bb components. The available dynamical, photometric and spectroscopic data are consistent with a binary
system with Ba being slightly (10-20\%) more massive than Bb. We obtain new orbital parameters which are in good
agreement with those in the literature.
\end{abstract}

\begin{keywords}
Instrumentation: high angular resolution - Instrumentation: adaptive optics - Stars: low-mass, brown
dwarfs - Binaries: close - Stars:individual: GJ569
\end{keywords}

%
\section{Introduction}
%

The lucky imaging (LI) technique proposed by \citet{FriedD:probgl} attracted the attention by professional astronomers
once low read-out noise detectors became available \citep[e.g.][]{BaldwinJ:diffl8, TubbsR:difflc, LawN:luckih}. Recently
it has been realized that the combination of LI and Adaptive Optics (AO) can benefit mutually and provide
high-resolution imaging close to the diffraction limit at optical and near-infrared (NIR) wavelengths
\citep[e.g. see][]{GladyszS:luckis,LawN:GettLAO,KervellaP:cloce}. This can be obtained by using a larger fraction of
data in the LI selection and/or keeping images with a better Strehl ratio than in conventional LI observations.

In this paper we present the results of combining the LI and AO techniques to produce high-angular resolution and
high-contrast imaging in the optical of the multiple system GJ569Bab which is a benchmark in the study and
characterization of brown dwarfs (BDs). GJ569A is an M2.5V cromospherically active star lying at a distance of 9.6-9.8
pc \citep{PerrymanM:HIPPARCOS,vanLeeuwenF:hippnr}. \citet{ForrestW:possbd} identified a faint companion to GJ569 and
argued the potential brown dwarf nature of such companion.  Using Keck AO observations \citet{MartinE:discv} resolved
GJ569B as a binary brown dwarf system with a separation of $\sim 0\farcs1$, a total mass of the system in the range
0.09-0.15~\MS, an age in the range 0.12-1.0~Gyr and an orbital period $\sim 3$ yr.  Further AO-based observations with
the Keck telescope \citep[e.g. ][]{LaneB:orbbd, ZapateroM:dynmbb, SimonM:gl569ms, KonopackyQ:highpd} and the Subaru and
HST telescopes \citep{ZapateroM:lithd, MartinE:reshs} have allowed precise determination of the dynamical masses and
orbital parameters of the binary system GJ569Bab (see Table~\ref{tab:orbit} in this paper) as well as a precise
determination of the spectral types of the GJ569B components: M8.5-9V and M9V for the Ba and Bb components, respectively
\citep{LaneB:orbbd,MartinE:reshs}.

For BDs an estimate of the mass is essential to determine their properties and evolution. A way to achieve a direct
measurement of masses is to observe close binary systems where the short orbital period allows for a complete sampling
of the orbit and from here a precise determination of the dynamical masses of the pair. Up to now a complete
characterization of the orbital motion has been achieved for a few BD binaries \citep[e.g. see][]{DupuyT:dynams,
  KonopackyQ:highpd}. In this context GJ569Bab constitutes a unique laboratory where to test the stellar evolutionary
models as it is among the shortest known period BD binary system. This has allowed to determine the orbit of the system
over several periods and from here a precise determination of its dynamical mass.

 The high angular resolution requested to spatially resolve faint systems like GJ569B into its components has been so far achieved
 with 10-m class telescopes and AO in the NIR or with the HST.. Our motivation to perform observations of GJ569 with
 LI+AO on a 4-m class telescope was twofold. First, to test the potential of this technique for high-angular resolution
 and high-contrast imaging in the optical regime. Second, to shed light on whether GJ569B is actually a triple system as
 suggested in some works in the literature \citep{MartinE:discv,KenworthyM:Gl569Bay,SimonM:gl569ms}.

In Section~\ref{Sect:Setup} we briefly describe the instrumental set-up.  Section~\ref{Sect:Observations} describes the
observations. Section~\ref{Sect:Analysis} reports the data calibration, reduction and analysis.
Section~\ref{Sect:Results} focuses on the photometry and astrometry of the GJ569 system components and discussion of the
results. We provide our conclusions in Sect.~\ref{Sect:Conclusions}.

%
\section{Instrumental setup.}\label{Sect:Setup}
%

FastCam is a LI-based instrument developed jointly by the Instituto de Astrof{\'\i}sica de Canarias (IAC) and
Universidad Polit{\'e}cnica de Cartagena (UPCT). In brief, FastCam consists of a commercial Electro Multiplying Charge
Coupled Device (EMCCD), a very simple versatile optics setup offering different plate scales to be accommodated at
different telescopes configurations and an FPGA-based\footnote{FPGA, Field Programmable Gate Arrays} on-line processing
and acquisition hardware. Its high degree of versatility has allowed us to collect images on a wide range of telescopes
with diameters from 1.5 to 4.2 m. Further details of the FastCam instrument and results achieved so far at different
telescopes are presented in \citet{OscozA:FastCam,LabadieL:highsr}.

The observations reported in this paper were acquired at the 4.2 meter William Herschel Telescope (WHT) at Observatorio
del Roque de Los Muchachos (ORM, La Palma). FastCam was installed at NAOMI, the AO facility at WHT.  It suffices here to
say it is an AO system based on a $8\times8$ subaperture Shack-Hartmann sensor and the most prominent feature of the
system is the use of a fully segmented deformable mirror. Details for NAOMI can be found in \citet{MyersR:naoao,
  BennC:naoosp}. We also resorted to non-AO LI observations with FastCam at the Cassegrain focus of the Nordic Optical
Telescope (NOT) at the ORM mainly for plate scale and orientation calibration.

The standard data acquisition of FastCam is that raw data are composed of cubes of 1000 images each. The full detector
of FastCam is an E2V EMCCD with 512$\times$512 pixels (commercial camera model Andor iXon DU-987) exhibits a maximum
frame rate of 35 Hz, although the exposure time for each individual frame can be set at values shorter than 1/35 seconds
but at the cost of incurring on overheads.

%
\section{Observations.}\label{Sect:Observations}
%

\begin{figure*}
  \centering
  \includegraphics[width=15.5cm]{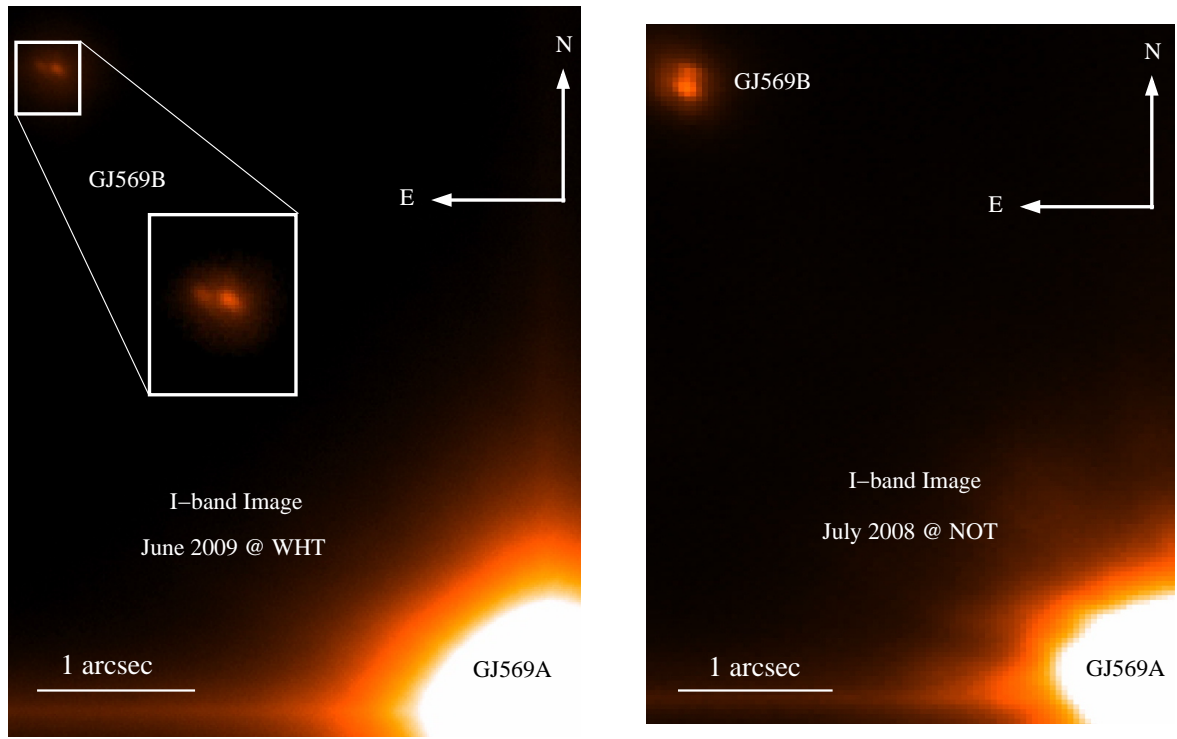}
  \caption{Left: I-band image of GJ569 observed on the 4th of June 2009 with FastCam at the science focus of the WHT
    telescope AO system. Right: GJ569 in the I-band observed on the 24th of July 2008 with the NOT telescope.  The image
    at the WHT is clearly resolved into the GJ569Ba and GJ569Bb components, while the NOT image acquired nearly 1
    year before suggests a binary system because of the elongation of GJ569B. Both images have been rotated by their
    respective Position Angle Offsets (see Section~\ref{SubSect:PlateCalib} to have the North and East axis pointing
    upwards and to the left, respectively).}
  \label{fig:GJ569image}
\end{figure*}

\subsection{AO+LI observations  with the WHT telescope.}\label{Subsect:ObsWHT}

On the 4th of June 2009 FastCam was configured for an individual frame exposure time of 30 msec and with optics
providing a field of view for FastCam of 7$\times$7 arcsec$^2$ (see Section~\ref{SubSect:PlateCalibGJ}). The primary
star GJ569A was used as the guide star for both the NAOMI and the FastCam real-time selection algorithm. Such real-time
selection is performed by the FPGA-based acquisition system and it is mostly used to detect any anomalies or
malfunctioning of the system by displaying the best 10\% frames based on the brightest pixel selection although the
entire frame set is kept for off-line processing. With this FastCam configuration, we collected 50 cubes which amount to
a total exposure time of 1500 seconds in the standard I Johnson-Bessell band (872.2~nm central wavelength, 298~nm FWHM)
and 824~nm central wavelength, 175~nm FWHM when the EMCCD QE curve is considered. Taking into account overheads, the
total time spent on GJ569 was 2340 seconds. During most of the observation time on GJ569 we had access to the
RoboDIMM\footnote{\url{http://www.ing.iac.es/Astronomy/development/seeing/Correl_DIMM.html}} seeing monitor which
reported an average seeing in the visible of $0\farcs49 \pm 0\farcs08$ (minimum seeing 0\farcs38, maximum seeing
0\farcs77) during the observation of GJ569. We also observed astrometric binaries for plate scale and orientation
calibration (see Section~\ref{SubSect:PlateCalibWDS}).

\subsection{LI observations  with the NOT telescope.}\label{Subsect:ObsNOT}

On the 24th July 2008 FastCam was configured for an individual frame exposure time of 20 msec and with optics providing
a field of view for FastCam of 16$\times$16 arcsec$^2$ (see Section~\ref{SubSect:PlateCalibNOT}). In the I
Johnson-Bessell band (same filter as with the WHT, see Section~\ref{Subsect:ObsWHT}) we collected 10 cubes which
correspond to a total exposure time of 200 seconds. No independent measurements for the seeing conditions were available
at the time although the combined image of all 10000 individual frames yields an estimate for the seeing of $\sim
0\farcs5$ at the detector plane. The primary star GJ569A was used as the guide star for the FastCam real-time selection
algorithm. We also observed the M15 globular cluster for plate scale and orientation calibration (see
Section~\ref{SubSect:PlateCalibNOT}).

\subsection{Observations  with the IAC80 telescope.}\label{Subsect:ObsIAC80}

For photometric calibration purposes we obtained $I$-band images of GJ569 on April 6 2010 using the CAMELOT instrument
mounted on the IAC80 telescope with an identical filter to the ones used in Sections~\ref{Subsect:ObsWHT} and
~\ref{Subsect:ObsNOT}. CAMELOT consists of a 2K$\times$2K CCD detector with a 0.304"/pix plate scale providing a field
of view of 10.4$\times$10.4~arcmin$^2$. We observed a series of 10 images with individual exposure time of 5s. Raw data
were reduced using routines within the IRAF environment \footnote{IRAF is distributed by National Optical Astronomy
  Observatories, which is operated by the Association of Universities for Research in Astronomy, Inc., under contract to
  the National Science Foundation.}. Images bias was subtracted using the overscan region and zero exposure time
images. The flat field correction made use of dome flats. Aperture photometry was performed using routines within the
DAOPHOT package. Weather conditions during our observations were photometric and average seeing ranged from 1.3 to
2". Instrumental magnitudes were transformed into apparent magnitudes in the Johnson-Bessel system using photometric
standard stars from \citet{LandoltA:UBVRIpss} obtained at different airmasses before and after our science target. Since
images of GJ569 were defocused to avoid saturation, aperture correction was applied to the photometry of the standards
to match the larger aperture used for our object. From these data, we have measured that the $I$-band magnitude of
GJ569A is 7.78$\pm$0.03.

%
\section{Data reduction and analysis}\label{Sect:Analysis}
%
The image selection employed is the same as with non-AO observations \citep{OscozA:FastCam,LabadieL:highsr} which
selects the best images as those exhibiting the largest ratio between the brightest pixel in the image and the rms of
the overall image. The images are sorted out from best to worst image according to the brightest pixel criterion with a
very small fraction of individual frames exhibiting saturated pixels (due either to cosmic rays or spurious electronic
events) removed. Then the user defines the percentage of best images to be kept on which a shift-and-add algorithm is
applied. Standard image reduction considered only bias subtraction as experience with the FastCam EMCCD has revealed its
excellent cosmetic and no need for flat fielding. Bias subs-traction, image selection and shift-and-add algorithm are
performed using our own software.

\subsection{Plate Scale and Orientation Calibration}\label{SubSect:PlateCalib}

\subsubsection{NOT calibration with M15}\label{SubSect:PlateCalibNOT}
As part of the observation program with the NOT telescope on the night of 24th July 2008, we also conducted observations
of the M15 globular cluster which, besides its specific science case, constitutes a very precise astrometric calibrator
yielding a plate scale accuracy of a few tens of microarcseconds and plate orientation below 0\fdg1 when the FastCam
image of the M15 core is compared against HST WFPC2 images of M15 \citep{vanderMarel:hsteimbh}. For the instrumental
setup on the 24th of July 2008 we measured a plate scale of $31.176 \pm 0.030$ mas and a Position Angle (P.A.) offset of
$-89.84\degr \pm 0.05\degr$ (i.e. North axis is at 89.84\degr clockwise from CCD y-axis).

\subsubsection{WHT calibration with WDS binaries.}\label{SubSect:PlateCalibWDS}
Observations of the M15 globular cluster were not possible on June 2009 as none of the cluster stars is bright enough for
efficiently locking the NAOMI loop. Because of this we observed four astrometric binaries (HD98231, HD105824, HD186858,
HD197913) selected from the Washington Double Star (WDS) Catalog \citep{WycoffG:datmd} with an adequate component
separation and brightness to act as astrometric calibrators. The calibration achieved from the WDS binaries (HD98231,
HD105824, HD186858, HD197913) yields a plate scale of $13.63 \pm 0.23$ mas and a Position Angle Offset of $91.3\degr \pm
0.9\degr$. The uncertainties assigned from the WDS astrometry corresponds to the dispersion of the values from each of
the WDS binaries.

\subsubsection{WHT calibration with GJ569.}\label{SubSect:PlateCalibGJ}

We are able to compute the GJ569B photocentre position with respect to GJ569A with high accuracy at the NOT image on
July 24th 2008 (see Table~\ref{tab:GJ569data2}). The orbital period of GJ569B around GJ569A is known a priori to be very
long. With the information in Table~\ref{tab:GJ569data2} we predict the coordinates of GJ569B with respect to GJ569A on
June 4th 2009 (WHT observation) implicitly dismissing GJ569A-B orbital displacements between the images at the NOT and
at the WHT nearly one year later.  From here a plate scale of $13.40 \pm 0.14$ mas and a Position Angle Offset of
$91.8\degr \pm 0.6\degr$ is derived.  A weighted average of these values with those obtained in
Section~\ref{SubSect:PlateCalibWDS} yields the calibration assumed in this work for the WHT data: $13.46 \pm 0.12$ mas
and $91.6\degr \pm 0.5\degr$ for the plate scale and Position Angle, respectively.

\subsection{PSF of the observations.}\label{SubSect:Results:AngularResolution}

\begin{figure}
  \centering
  \includegraphics[angle=90,width=9cm]{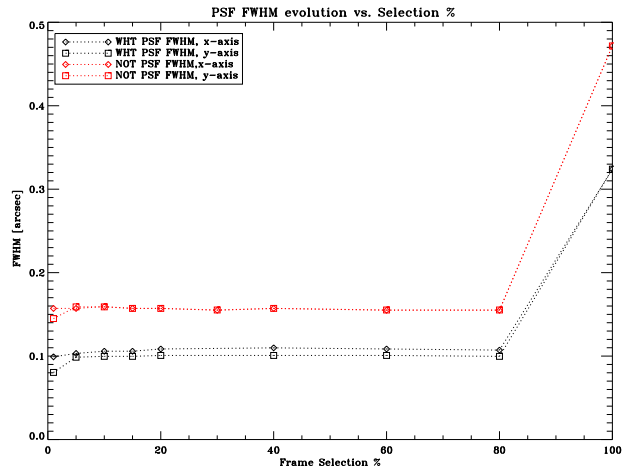}
    \caption{Evolution of the PSF FWHM at NOT (red) and WHT(black) as a function of the percentage of images kept in the
      selection procedure. The PSF is directly the image of GJ569A. In this plot the x-axis corresponds to the direction
      at which the PSF is widest and the y-axis is perpendicular to the x-axis. Notice these x and y axes do not
      necessarily correspond to the CCD x and y axes. The asymmetry of the PSF for the WHT observations is believed to
      be caused by the absence of an atmospheric dispersion corrector.}
   \label{fig:FWHMevo}
\end{figure}

\begin{figure*}
  \centering
  \includegraphics[angle=90,width=14cm]{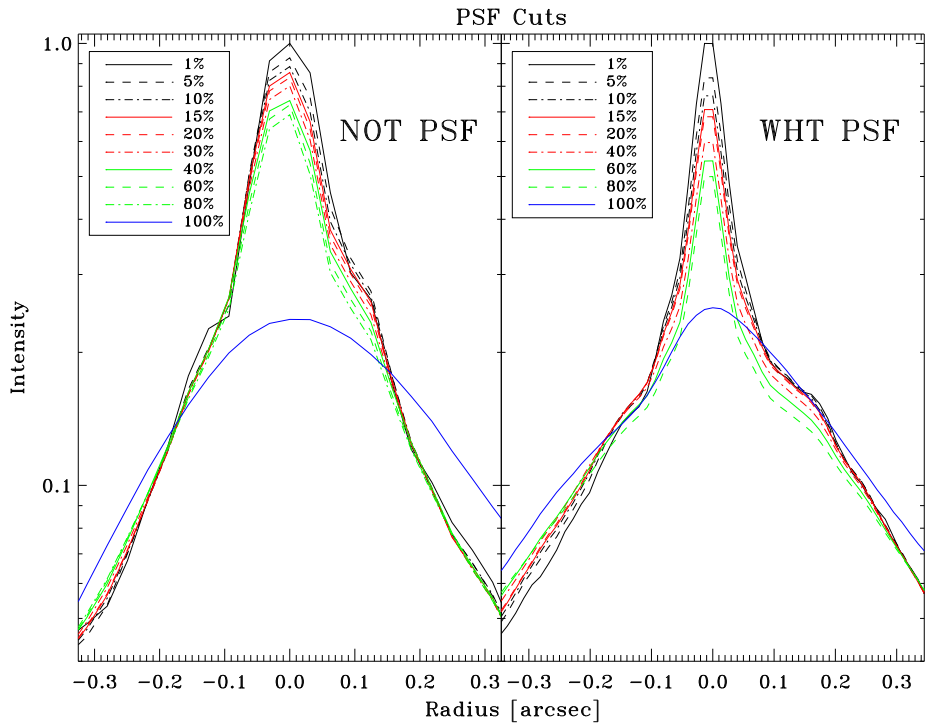}
    \caption{Cuts of the PSFs at the NOT and WHT as a function of the percentage of images kept in the selection
      procedure. The PSF is directly the image of GJ569A. The PSFs have been normalized so that the highest PSF exhibits
      peak unity and all of them have the same volume under the PSF.The gain in resolution from the WHT with respect to
      the NOT is due both to the increasing WHT diameter plus the use of the NAOMI system.Notice the blue line in the
      NOT plot represents the seeing-limited PSF while in the case of the the blue PSF in the WHT diagram corresponds to
      a partially tip-tilt removed by the NAOMI system. }
   \label{fig:PSFcuts}
\end{figure*}

The NAOMI system is optimized to deliver well corrected wavefronts above $1.0~\mu$m, while in the optical the aim is not
to deliver well corrected images (i.e. with a significant Strehl ratio) but simply to achieve a reduction by a factor 2
of the FHWM with respect to the seeing-limited images. The combination of AO+LI allows to achieve FHWM reductions by at
least a factor 5 with respect to the seeing-limited images as shown in Fig.~\ref{fig:FWHMevo} and Fig.~\ref{fig:PSFcuts}
where we show the evolution of the FWHM and different cuts of the PSF versus the LI \% selection, respectively. The PSF
cuts shown in Fig.~\ref{fig:PSFcuts} correspond to the PSF as measured on GJ569A for both the NOT and the WHT images.

The FWHM evolution depicted in Fig.~\ref{fig:FWHMevo} shows the usual low sensitivity of the FWHM upon LI \%, a trend
already pointed out in previous works \citep[e.g. see][]{LawN:GettLAO}. To understand this trend one must take into
account that most of the turbulence is concentrated in the tip-tilt modes which are removed in the LI scheme, while
higher-order turbulent modes tend to distort the PSF but without enlarging it significantly. The effect of these
higher-order modes in the PSF can be understood as a leak of power towards outside the PSF core, giving rise to the
familiar LI PSFs consisting on a very well defined narrow core on top of a much wider and swallower halo. Qualitatively
one can think that the better the AO correction or the most stringent the LI frame selection are, the more prominent the
core over the halo becomes. This is in fact what we observe in the PSF cuts in Fig.~\ref{fig:PSFcuts}, where the PSFs
have been normalized to have unit volume. In this way we notice that when we restrict ourselves to very low LI \%
selection the PSF peaks much higher with respect to the halo that when using a larger LI \% selection. This leads
directly to a trade-off when having to decide the frame \% to keep when one also considers the desirable SNR to
achieve. 

\subsection{The PSF fitting technique}\label{SubSect:LMPSFfitting}

In order to perform an estimate of the relative photometry and astrometry between the primary component and each of the
resolved components of GJ569B we have resorted to a Levenberg-Marquadt (LM) non-linear least-squares
\citep{BevingtonP:datare} PSF fitting algorithm\footnote{Using the MPFIT toolset by C. Markwardt available at
  \url{http://cow.physics.wisc.edu/~craigm/idl/idl.html}}. We assume the PSF measured at the GJ569A location and the PSF
at GJ569B position do not differ significantly. This is justified by the excellent seeing conditions experienced during
the observations and the fact that the isoplanatic angle at the ORM  at 820 nm is frequently larger than the 5
arcsec separation between GJ569A and B \citep[see e.g.][]{FuensalidaJ:staap,LawN:luckih}. It has also been pointed out
by \cite{LawN:luckih} that as the LI technique selects the best fraction of frames where the turbulence is smallest, it
is expected that those frames also exhibit larger isoplanatic angles than would be observed in traditional long exposure
imaging. Therefore, for a separation of about 5\arcsec we should be within the isoplanatic angle at $\sim 820$~nm and
the assumption of uniformity of the PSF at the GJ569A and GJ569B locations is well justified.

Our fitting model is applied to a relatively small box with only GJ569B and assumes the signal on each pixel in the box,
$f(x,y)$, is due to two components GJ569Ba and G569Bb located at positions $(x_a,y_a)$ and $(x_b,y_b)$, respectively,
and with fluxes $a$ and $b$ relative to the flux of GJ569A, plus a possible offset ($\mathrm{OFF}$) associated either to
sky background or a bias offset in the final image:

 \begin{equation}
f(x,y)= a \cdot PSF(x-x_a,y-y_a) + b\cdot PSF(x-x_b,y-y_b) + \mathrm{\mathbf{OFF}}
\end{equation}

Note that in our LM PSF fitting scheme we do not fit an analytical PSF function nor attempt first a fit to the PSF
measured at GJ569A. Instead we use directly the measured GJ569A PSF in the WHT image which is moved across a $100\times
100$ pixel box by performing FFT-based shifts. In this way, the PSF fitting technique also provides us with a
measurement of the distance between each of the GJ569B components with respect to each other and with respect to GJ569A.

\begin{figure*}
  \vspace{-1cm}
  \centering
  \resizebox{\hsize}{!}{\includegraphics{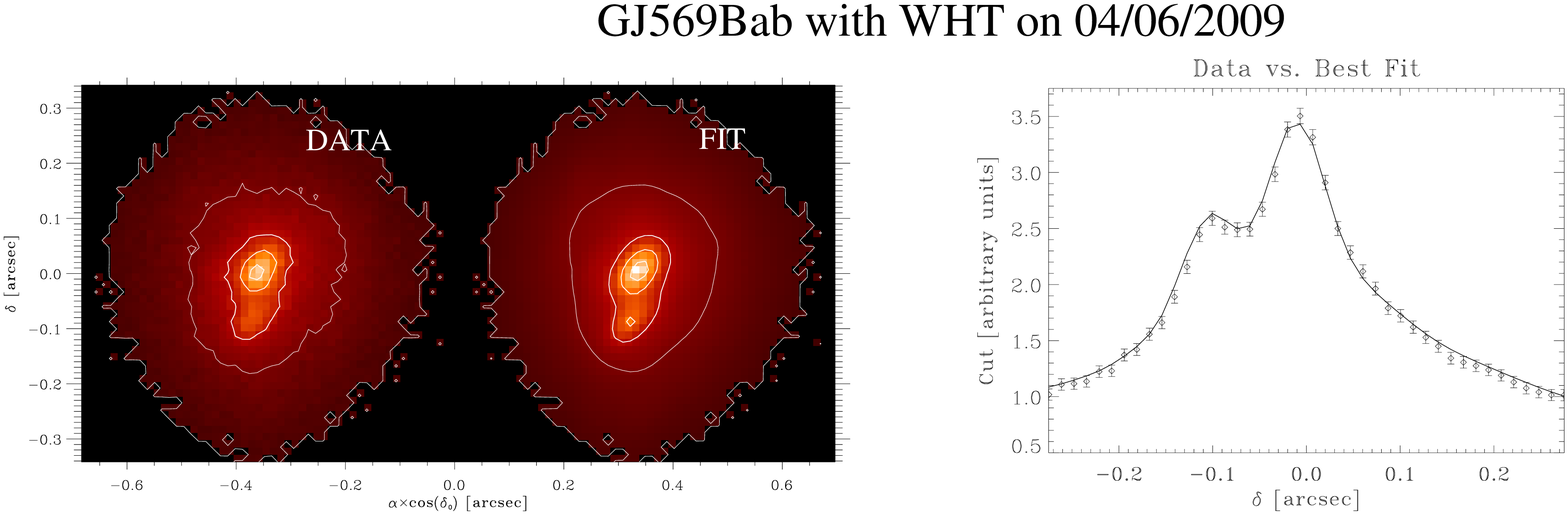}}
  \caption{Comparison of actual images to best fit models and respective cuts along lines joining GJ569Ba to GJ569Bb for
    data acquired with the WHT on 4/6/2009. The oversampling of the WHT data and a better PSF due to the use of the AO
    facility at the WHT allows to resolve the binary nature of GJ569B allowing the LM PSF fitting technique to achieve a
    good fit to the data. On both images dark areas indicate regions whose pixels are not sufficiently above the
    background threshold to be taken into consideration in the PSF fitting (e.g. $\eta$ parameter in
    Tables~\ref{tab:LMfit} and \ref{tab:LMfit_LIf}). No rotation to align the North and East axis with the usual y and
    negative axes has been applied to the image.}

  \label{fig:PSFfit}
\end{figure*}

The results of the LM PSF fitting technique on the image composed by the best 16\% of the frames from the WHT
observation are summarized in Table~\ref{tab:LMfit} for different values of the threshold background parameter
($\eta$). This parameter determines the number of pixels within the $100\times 100$-pixel box which are finally
considered in the fitting process: only those pixels exceeding the background level by a factor $\eta$ the rms
background. Pixels below such threshold are shown in black in Fig.~\ref{fig:PSFfit}. As it can be seen in
Table~\ref{tab:LMfit}, when applied to the WHT image the technique is rather insensitive to the choice of pixels in the
fitting (i.e. little sensitivity to the $\eta$ parameter) and provides the relative photometry with excellent accuracy.

\begin{table*}
\centering
\caption{LM PSF fitting results on the WHT image from best 16\% of frames.}
\label{tab:LMfit}
\begin{tabular}{c c c c c c c}
\hline \hline
$\eta$&   $a\times 10^3$  &   $b\times 10^3$  &      $b/a$        &  $\rho (mas)$  &  P.A. (\degr)  & $\chi^2/\nu$ \\ \hline
 2.0  & $3.743 \pm 0.023$ & $2.126 \pm 0.022$ & $0.568 \pm 0.009$ & $98.6 \pm 1.1$ & $78.0 \pm 0.7$ &   0.92     \\
 2.5  & $3.733 \pm 0.024$ & $2.113 \pm 0.023$ & $0.566 \pm 0.009$ & $98.6 \pm 1.1$ & $78.0 \pm 0.7$ &   0.95     \\
\textbf{3.0} & $\mathbf{3.715 \pm 0.024}$ & $\mathbf{2.096 \pm 0.023}$ & $\mathbf{0.564 \pm 0.009}$ & $\mathbf{98.4 \pm
  1.1}$ & $\mathbf{78.0 \pm 0.7}$ & \textbf{1.04} \\
 3.5  & $3.689 \pm 0.024$ & $2.076 \pm 0.023$ & $0.563 \pm 0.009$ & $98.2 \pm 1.1$ & $78.0 \pm 0.7$ &   1.10     \\
 4.0  & $3.666 \pm 0.025$ & $2.054 \pm 0.024$ & $0.560 \pm 0.009$ & $97.8 \pm 1.1$ & $78.1 \pm 0.7$ &   1.16     \\
 5.0  & $3.626 \pm 0.026$ & $1.994 \pm 0.025$ & $0.550 \pm 0.009$ & $97.5 \pm 1.1$ & $78.2 \pm 0.7$ &   1.27     \\ 
 6.0  & $3.561 \pm 0.028$ & $1.920 \pm 0.026$ & $0.539 \pm 0.010$ & $96.8 \pm 1.1$ & $78.1 \pm 0.7$ &   1.33     \\ 
 7.0  & $3.490 \pm 0.029$ & $1.821 \pm 0.028$ & $0.522 \pm 0.010$ & $95.8 \pm 1.2$ & $78.1 \pm 0.7$ &   1.25     \\ 
\hline
\end{tabular} 
\end{table*}

For the PSF fitting model to work properly and provide with meaningful error bars, the pixels in the image have to be
assigned realistic error bars. During the analysis of the data the simple choice whether a given pixel was background
limited or source limited was not good enough.  At this point it revealed particularly useful the fact that the LI
scheme allows for a precise distribution of values of the signal for each pixel in the final image since we have access
to all the individual frames. For the final WHT image made of the best 16\% individual frames, the value assigned to
each pixel is the mean of a distribution sampled with 8000 values and we also generated an error map computed as the
error of the mean.

\begin{table*}
\centering
\caption{LM PSF fitting results as a function of the percentage of frames kept to generate the WHT image.}
\label{tab:LMfit_LIf}
\begin{tabular}{c c c c c c c}
\hline \hline
LI \% &   $a\times 10^3$  &   $b\times 10^3$  &      $b/a$        &  $\rho (mas)$   &  P.A. (\degr)  & $\chi^2/\nu$ \\ \hline
  1   & $3.35  \pm 0.06 $ & $1.77  \pm 0.05 $ & $0.528 \pm 0.021$ & $ 96.5 \pm 1.8$ & $77.8 \pm 1.1$ &   0.880    \\
  5   & $3.59  \pm 0.05 $ & $2.07  \pm 0.04 $ & $0.576 \pm 0.018$ & $ 98.1 \pm 1.5$ & $78.2 \pm 0.9$ &   0.360    \\
 10   & $3.688 \pm 0.029$ & $2.082 \pm 0.026$ & $0.565 \pm 0.010$ & $ 98.5 \pm 1.2$ & $78.2 \pm 0.7$ &   0.722    \\

\textbf{16} & $\mathbf{3.715 \pm 0.024}$ & $\mathbf{2.096 \pm 0.023}$ & $\mathbf{0.564 \pm 0.009}$ & $\mathbf{98.4 \pm
  1.1}$ & $\mathbf{78.2 \pm 0.6}$ & \textbf{1.04} \\

 20   & $3.710 \pm 0.022$ & $2.125 \pm 0.021$ & $0.573 \pm 0.008$ & $ 98.3 \pm 1.1$ & $78.0 \pm 0.6 $ &   1.27     \\
 40   & $3.745 \pm 0.017$ & $2.148 \pm 0.016$ & $0.574 \pm 0.006$ & $ 98.3 \pm 1.0$ & $78.3 \pm 0.6 $ &   2.11     \\
 60   & $3.765 \pm 0.015$ & $2.140 \pm 0.014$ & $0.569 \pm 0.005$ & $ 98.9 \pm 1.0$ & $78.2 \pm 0.6 $ &   2.77     \\ 
 80   & $3.755 \pm 0.014$ & $2.132 \pm 0.013$ & $0.568 \pm 0.005$ & $ 98.6 \pm 1.0$ & $78.1 \pm 0.5 $ &   3.24     \\ \hline
\end{tabular} 
\end{table*}

The LM PSF fitting technique was also robust versus the LI selection fraction, exhibiting little dispersion on the
derived parameters as it is shown in Table~\ref{tab:LMfit_LIf} for the choice $\eta=3.0$. The choice of the 16\%
fraction to generate the final LI image for the WHT data was based entirely on the value of the reduced $\chi^{2}$
achieved in this series of fits: the WHT image generated with a 16\% of the best images exhibits $\chi^{2}/\nu=1.04$
($\nu$ being the number of degrees of freedom).

The case for the NOT image is less favourable. When applied to the NOT image the LM PSF fitting was not that robust
and we believe this is due both to the lower number of images under consideration but mainly to the fact the blending of
both sources hampers a good fit.

\subsection{Detectability curves.}\label{SubSect:Results:contrast}

\begin{figure*}
  \centering
  \includegraphics[angle=90,width=15.5cm]{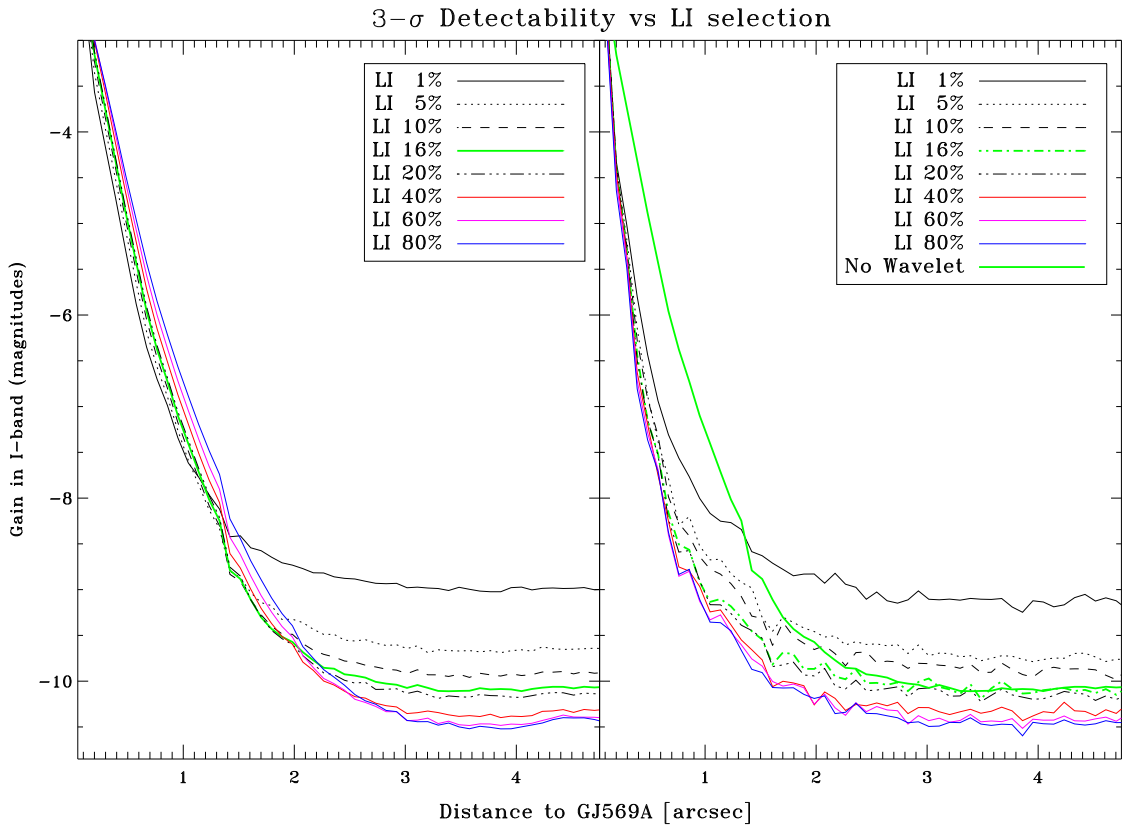}
    \caption{ Evolution of the 3$\sigma$ detectability versus the percentage of images kept to generate the LI
      image. The left panel shows the 3$\sigma$ detectability curves obtained directly from the image. Right panel
      shows the 3-$\sigma$ detectability curve upon filtering with wavelets (see main text). The green thick solid lines
    in both plots correspond to a selection of the best 16 \% frames without wavelet filtering and is displayed in the
    right panel for comparison purposes. Notice the big improvement by the wavelet filtering at angular distances below
    1\farcs5, a slight gain at angular distances in the range [1\farcs5, 2\arcsec] and no net gain beyond 2\arcsec.}
   \label{fig:ContrastWHT}
\end{figure*}

A key point to characterize the combination of AO-assisted LI systems for the detection of low-mass companions is the
study of the detectability curves as a function of the distance to the star. This has been done in
Fig.~\ref{fig:ContrastWHT} where on the left panel we show the $3\sigma$ detectability curves for a variety of
percentages of the images kept to generate the final images. On the right panel we have plotted the detectability curves
on the same images upon filtering them with a wavelet transform algorithm. The details of this wavelet-based
post-processing algorithm are described in \citet{LabadieL:imprhc} and the general purpose of such filtering technique
is the increase in the contrast ratio achievable by suppressing the low-frequency halo of the primary star which hides
the presence of much fainter companions \citep{MasciadriE:exoruw}. Both sets of detectability curves were obtained as
follows: at a given distance $\rho$ from the primary star we identify all possible sets of small boxes of size the FWHM
of the PSF (i.e. $7 \times 7$ pixel boxes). Only regions of the image showing structures easily recognizable as spikes
due to diffraction of the telescope spider and/or artifacts on the read-out of the detector are not considered. For each
of the valid boxes on the arc at distance $\rho$ the standard deviation of the image pixels within the $7\time 7$-pixel
boxes is computed. The value assigned to the $3\sigma$ detectability curve at distance $\rho$ is 3 times the mean value
from the standard deviations of all the eligible boxes at distance $\rho$ from GJ569A. From the curves in both panels in
Fig.~\ref{fig:ContrastWHT} we notice the following behaviours:

\begin{enumerate}
\item On the non-wavelet filtered image, at short distances (< 1\farcs5) from the primary star, the lower the fraction
  of data used the higher the detectability gain is with a maximum gain of around $\sim 0.9$ magnitudes between the 80\%
  and 1\% frame selection. This improvement in contrast is due a slightly better angular resolution with a very
  restrictive frame selection. As we move away from the primary star, the detectability curves approach to each other
  until eventually the situation is reversed. At large distances the image is dominated by background/detector noise so
  that the higher the percentage of frames used to generate the image, the deeper the detectability curve gets.

\item On the wavelet-processed image there is no transition between the inner region (i.e separation from primary below
  1\farcs5) and the outer region. The detectability at the inner regions is always improved with respect to the
  non-processed image and, in addition, the higher the percentage of data kept for the image, the deeper the
  detectability gain. Above 1\farcs5 separation from the primary star, there is no real difference between the
  detectability achieved on the original and the wavelet-processed image, as it is expected at those regions of the
  image which are dominated by background/detector noise which can not be removed by the wavelet post-processing
  technique.
\end{enumerate}

\section{Results.}\label{Sect:Results}

\subsection{Photometry of the GJ569 system.}\label{SubSect:Results:photometry}
 
The final $I$-band differential magnitudes ($\Delta m_{AB}=m_{B} - m_A$) derived by the LM PSF fitting technique described in
Section~\ref{SubSect:LMPSFfitting} are summarized in Table~\ref{tab:photometry}. The magnitude difference between
GJ569Ba and GJ569Bb we find is very similar to the 0.7 mag found by \citet{MartinE:reshs} using the F814W filter in the
HST which resembles the $I$-band. These values  are slightly larger than the 0.5 magnitude difference
measured in the $J,H$ and $K$-band by \citet{MartinE:discv}. All these data are consistent with the spectral determination
of M8.5-M9 for the brown dwarf binary derived in the near-infrared by \citet{LaneB:orbbd}. However, given the
uncertainties, we can not discard that both components have a similar spectral type of M9, as obtained by
\citet{MartinE:reshs} using optical spectra.

\begin{table}
  \caption{Differential magnitudes in I band with FastCam at WHT}
  \label{tab:photometry}
  \centering
  \renewcommand{\footnoterule}{}  
  \begin{tabular}{c c}
    \hline \hline \\
    $\Delta m_{AB}$   &$5.590 \pm 0.004$ \\        
    $\Delta m_{ABa}$  &$6.075 \pm 0.007$ \\         
    $\Delta m_{ABb}$  &$6.697 \pm 0.012$  \\
    $\Delta m_{BaBb}$ &$0.622 \pm 0.017$  \\ \hline   
\end{tabular} 
\end{table}

Using the $I$-band photometry obtained in the IAC80 for GJ569A (see Section~\ref{Subsect:ObsIAC80}), and the
differential magnitude between the primary and the brown dwarf binary found with FastCam (see
Table~\ref{tab:photometry}), we derived an integrated photometry of $I$=13.37$\pm$0.03 for GJ569B with the WHT image and
$I$=13.41$\pm$0.03 with the NOT image. As remarked previously, the lack of enough spatial resolution on the NOT image
does not allow to resolve the GJ569B binary system, but aperture photometry of the whole GJ569B system is compatible
with what we derive from the LM PSF fitting of the WHT image. This value is not in good agreement with the previous
result by \citet{ForrestW:possbd} ($I$=13.88$\pm$0.2), probably affected by the contamination of the primary. To check a
possible variability of GJ569A , we have compared the apparent magnitude of this object and the G1V star HD130948, which
was observed during the same nights in the IAC80 and the NOT. We have found no relative difference in their $I$-band
magnitude at the level of a few tens of millimagnitudes. The absolute photometry of both components of the brown dwarf
binary was conducted in a similar way and a summary with the apparent magnitudes as derived from this work and in the
literature is given in Table~\ref{tab:photometry2}.

\begin{table*}
  \caption{Updated compilation of GJ569 photometry.}
  \label{tab:photometry2}
  \centering
  \renewcommand{\footnoterule}{}  
  \begin{tabular}{c c c c c c c}
    \hline \hline
            & Spec. Type  & V magnitude     &     I magnitude          &      J magnitude    &     H magnitude     & K magnitude          \\ \hline

    GJ569A  & M2.5V\refd  &$10.11\dagger$\refb & $ 7.78\pm 0.03$\refj  &$6.633\pm 0.023$\refe&$5.990\pm 0.021$\refe&$5.770\pm 0.018$\refe \\ \\

    GJ569B  &             &                 & $13.88\pm 0.2$\refa      &$10.61\pm 0.05$\refd &$10.16\pm 0.10$\refa &  $9.56\pm 0.10$\refa \\
            &             &                 & $13.37\pm 0.03^{*}$\refj &$10.75\pm 0.06$\refi &$10.15\pm 0.04$\refi&  $9.45\pm 0.05$\refd \\ 
            &             &                 & $13.41\pm 0.03^{**}$\refj&                     &                     & $9.62\pm 0.03$\refi \\ 

    GJ569Ba &M8.5-9V\refdf&                 & $13.86\pm 0.03$\refj     &$11.14\pm 0.07$\refd &$10.43\pm 0.04$\refg & $10.02\pm 0.08$\refd \\
            &             &                 &                          &$11.18\pm 0.08$\reff &                     &  $9.86\pm 0.10$\refg \\ \\

    GJ569Bb &  M9V\refdf  &                 & $14.48\pm 0.03$\refj     &$11.65\pm 0.07$\refd &$11.04\pm 0.05$\refg & $10.43\pm 0.08$\refd \\
            &             &                 &                          &$11.69\pm 0.08$\reff &                     & $10.39\pm 0.06$\refg \\ \\

    Bb/Ba   &             &                 &$0.70  \pm 0.23\ddagger$\reff& $0.5 \pm 0.2$\refc    & $0.5 \pm 0.1$\refc  & $0.5 \pm 0.1$\refc  \\
            &             &                 &$0.622 \pm 0.017$\refj    & $0.51 \pm 0.02$\refd & $0.61 \pm 0.03$\refg& $0.41 \pm 0.03$\refd \\ 
            &             &                 &                          &                      & $0.57 \pm 0.04$\refg& $0.41 \pm 0.13$\reff \\ 
            &             &                 &                          &                      &                     & $0.61 \pm 0.03$\refg \\  
            &             &                 &                          &                      &                     & $0.500\pm 0.008$\refh\\  \hline
\end{tabular} 
  \begin{list}{}{}

  \item[References:] \refa\citet{ForrestW:possbd}; \refb\citet{PerrymanM:HIPPARCOS}; \refc\citet{MartinE:discv};
    \refd\citet{LaneB:orbbd}; \refe\citet{CutriR:2MASS};\reff\citet{MartinE:reshs}; \refg\citet{SimonM:gl569ms};
    \refh\citet{KonopackyQ:highpd}; \refi\citet{DupuyT:studpd};\refj This work.

  \item[Notes:] $(\dagger)$10.11 V Johnson magnitude converted from Hipparcos photometric system $H_p = 10.201 \pm 0.004$;
    $^\mathrm{(*)}$Unresolved GJ569B $I$-band magnitude in June 2009 at WHT; $^\mathrm{(**)}$Unresolved GJ569B $I$-band
    magnitude in July 2008 at NOT; $(\ddagger)$Differential magnitude using the F814W filter in the HST which resembles the $I$-band.

\end{list}
\end{table*}

\begin{figure}
  \centering
  \includegraphics[width=8cm]{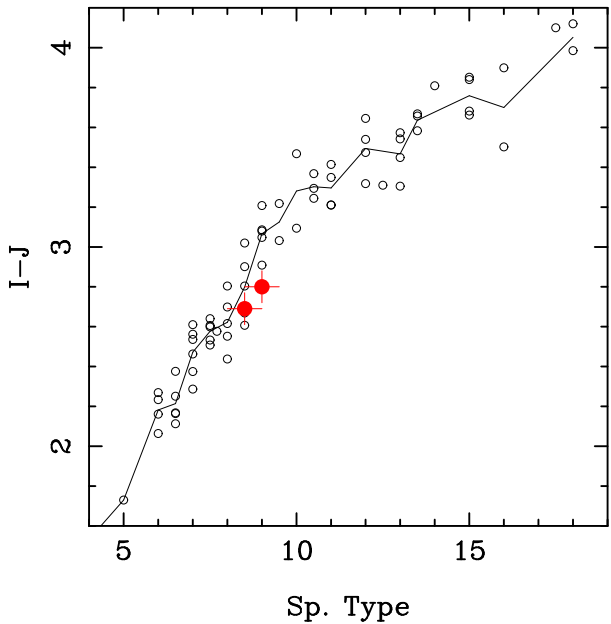}

    \caption{ $I-J$,spectral type diagram of GJ569Ba and GJ569b (solid red circles) in comparison with M (M5 to M9
      correspond to x-axis values 5 to 9) and L field dwarfs (L0 to L8 correspond to x-axis values 10 to 18) from
      \citet{LiebertJ:rip2m} (open circles).  The average color of these field dwarfs with the same spectral type are
      indicated by the solid line. The $J$-ban photometry obtained from \citet{LaneB:orbbd} in the CIT system has been
      transformed into the 2MASS system using the relations in \citet{CarpenterJ:colort2M}.}
   \label{fig:ColorIJ}
\end{figure}

With the $J$-band photometry in \citet{LaneB:orbbd} we derived a $I-J$ color of 2.72$\pm$0.08 and 2.83$\pm$0.08 for
GJ569Ba and GJ569Bb, respectively. In Fig.~\ref{fig:ColorIJ} we show the $I-J$ color of both components in comparison
with M and L field dwarfs of the same spectral type from \citet{LiebertJ:rip2m}. We have adopted the spectral
determination of M8.5-M9 for the brown dwarf binary derived in the near-infrared by \citet{LaneB:orbbd}. From this
figure we can see that our estimated $I-J$ colors are in good agreement with these spectral types.

According to the absolute magnitudes and colors of field dwarfs \citep{KirkpatrickJ:lowmc,LiebertJ:rip2m}, we expect a
relative magnitude difference of $\sim$ 0.4 mag in the $I$-band, and a $I-J$ color difference of 0.27$\pm$0.22 for a
M8.5/M9.  All these data are also consistent with previous determinations of spectral types, in particular for those
deriving a slightly earlier spectral type for GJ569Ba than for the Bb component.

\subsection{Astrometry of GJ569A-GJ569B.}\label{SubSect:Results:Astrometry}

\begin{figure}
  \centering
  \includegraphics[angle=90,width=10.3cm]{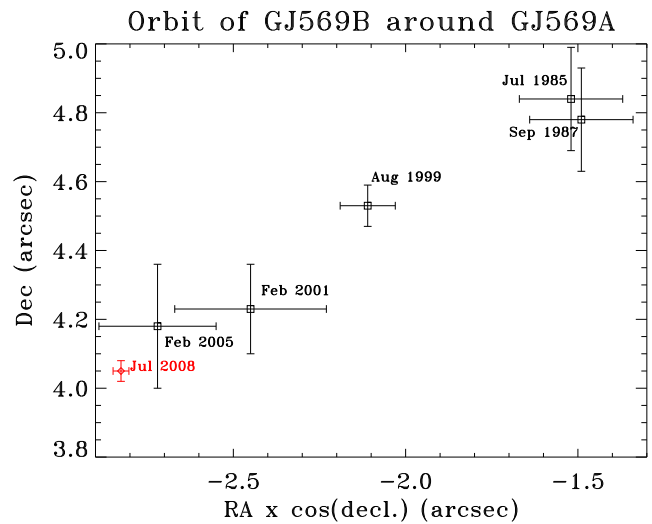}
    \caption{Compilation of positions of GJ569B on its long-period orbit around GJ569A using measurements in
      Table~\ref{tab:GJ569data2} and our point in red. GJ569A would be located at coordinates (0,0), which falls on
      purpose outside of the plot in order to make visible all points with their respective uncertainties.}
   \label{fig:GJ569data2}
\end{figure}

The compilation of all astrometric data on the GJ569B orbit around GJ569A, including our new measurement from the NOT
observation, is provided in Table~\ref{tab:GJ569data2} and shown in Fig.~\ref{fig:GJ569data2} with the value obtained
in this work in red. The higher accuracy of our measurement is immediately noticeable and it is due to the very precise
image statistics combined with the high angular resolution PSFs provided by the LI technique and a very accurate NOT
plate calibration, as discussed in Section~\ref{SubSect:PlateCalib}.

Only the photocentre positions in \citet{ForrestW:possbd} and in this work correspond to observations in the I band,
while the rest of the observations have been conducted in the NIR. Since the flux ratios between the GJ569B components
differ slightly from the NIR to the I-band measurement, we realize there might be a slight bias in the photocentre
determination. In principle such a bias could be avoided by a proper weighting with the individual masses of the GJ569B
pair, but due to the uncertainties in the individual GJ569Ba and GJ569Bb masses, the estimate of the barycenter remains
essentially the same as the computation of the photocentre. The limited coverage of the GJ569B orbit around GJ569A (see
Fig.~\ref{fig:GJ569data2}) does not allow for a reliable orbit computation but still we estimate a low eccentricity
long-period orbit of $\sim$400 years with a semi-major axis of 4\farcs3.

\begin{table}
\caption{Astrometry of the binary system GJ569A-GJ569B.} 
\label{tab:GJ569data2}
\centering
\renewcommand{\footnoterule}{}  
\begin{tabular}{c c c c }
\hline \hline
            &                 & Separation        &   PA     \\
Date        &      Ref.       & (arcsec)          &  (\degr) \\ \hline
1985 Jul 28 & \refa  & $5.07 \pm 0.15$   &  $17.4 \pm 1.7$   \\
1987 Sep 18 & \refa  & $5.01 \pm 0.15$   &  $17.3 \pm 1.7$   \\
1999 Aug 29 & \refb  & $5.00 \pm 0.05$   &  $25.0 \pm 0.9$   \\
2001 Feb  8 & \refc  & $4.89 \pm 0.04$   &  $30   \pm 3  $   \\
2005 Feb 25 & \refd  & $4.99 \pm 0.19$   &  $33.1 \pm 1.9$   \\
2008 Jul 24 & \refe  & $4.94 \pm 0.04$   &  $34.90\pm 0.05$  \\ \hline
\end{tabular} 
 \begin{list}{}{}
 \item[References:] \refa\citet{ForrestW:possbd}; \refb\citet{MartinE:discv}; \refc\citet{LaneB:orbbd};\refd\citet{SimonM:gl569ms}; \refe This work
 \end{list}
\end{table}

\subsection{Orbital analysis of the GJ569B system.}\label{SubSect:Results:orbit}

\begin{figure*}
  \centering
  \resizebox{\hsize}{!}{\includegraphics[angle=90,width=5cm]{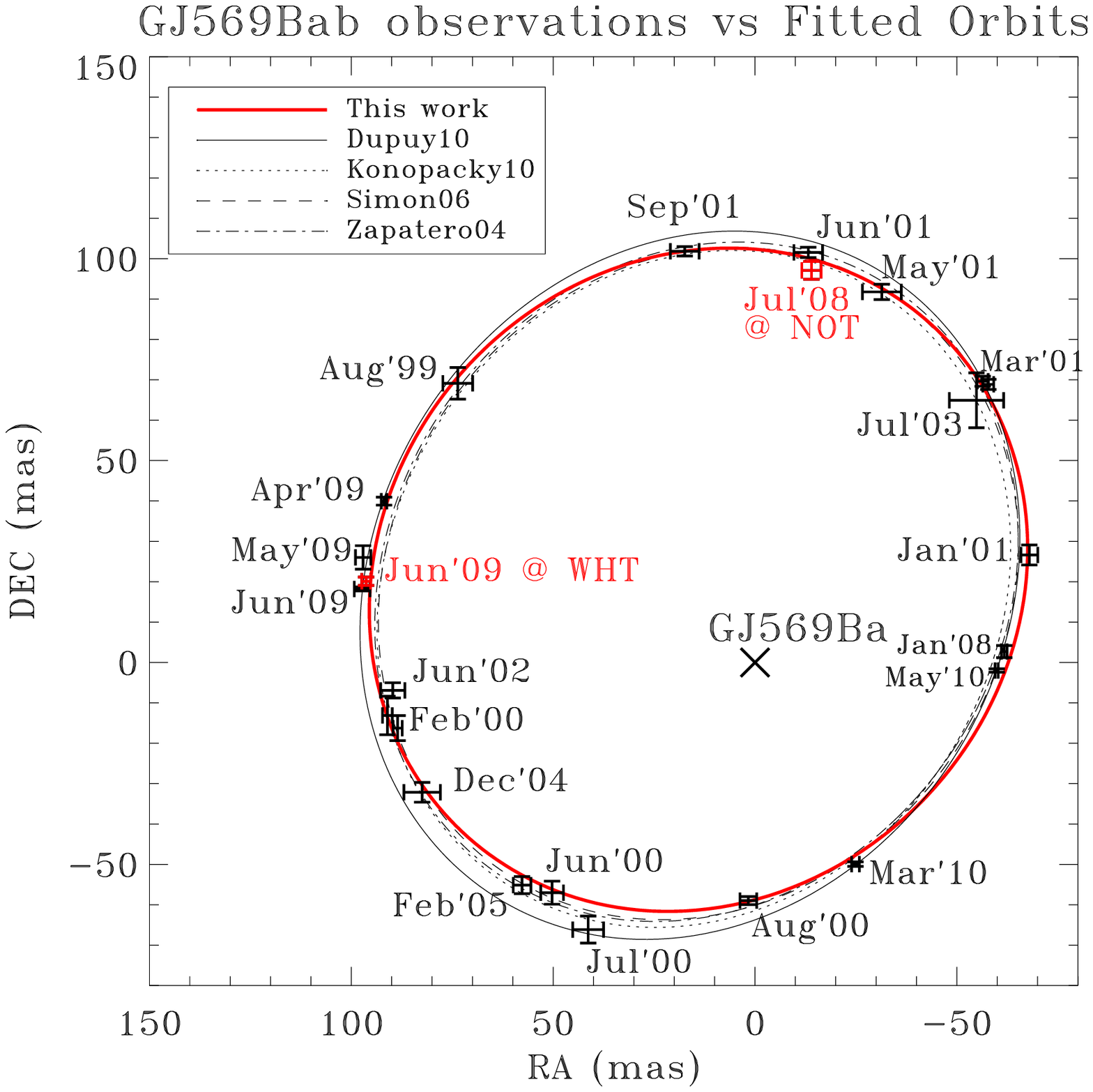}}
  \caption{Compilation of astrometric data on the GJ569Bab pair with our points in red (see Table~\ref{tab:GJ569data}).
    Overplotted there are the different best-fit Keplerian orbits for the GJ569 multiple system found in the literature
    and derived in this work and whose orbital parameters are summarized in Table~\ref{tab:orbit}.}
  \label{fig:DataOrbits}
\end{figure*}

As discussed in Section~\ref{SubSect:LMPSFfitting}, the LM PSF fitting procedure applied to the WHT images allows us to
obtain with high precision the relative position of Bb with respect to Ba (see Table~\ref{tab:GJ569data}). The same PSF
fitting approach applied to the unresolved GJ569B NOT image delivers a relative astrometry for these two components with
higher error bars. Fig.~\ref{fig:DataOrbits} summarizes all the observations in the literature with our measurements in
red. Our WHT astrometric value on June 2009 is nearly over imposed on the Keck AO-based observations by
\cite{DupuyT:studpd} and \citet{KonopackyQ:highpd}, the latter observved GJ569B within seven days after our
observation. Overplotted we show the different orbits found in the literature together with our best Keplerian orbit to
the same data plus our two new points. The orbital parameters of all these orbits are provided in Table~\ref{tab:orbit}
together with the total mass of the GJ569B system obtained by direct application of Kepler's third law. A very good
agreement is found for the orbital parameters as well as for the estimated total mass of the system except from the
value derived by \cite{DupuyT:studpd} (see discussion later in this section.)

\begin{table}
\caption{Astrometry of the binary system GJ569Ba-Bb.}
\label{tab:GJ569data}
\centering
\renewcommand{\footnoterule}{}  
\begin{tabular}{c c c c c c }
\hline \hline
             &         &  Teles-   &           &  Separation              &   PA               \\
Date         &   Ref.  &   cope    & In fit    &  (arcsec)                &  (\degr)           \\ \hline
1999 Aug 29  &  \refa  &   Keck    &   Yes     &  $0.101   \pm 0.001  $   &  $ 46.8 \pm 3  $   \\
2000 Feb 18  &  \refb  &   Keck    &   Yes     &  $0.092   \pm 0.001  $   &  $ 98.2 \pm 3  $   \\
2000 Feb 25  &  \refb  &   Keck    &   Yes     &  $0.090   \pm 0.001  $   &  $100.4 \pm 2  $   \\
2000 Jun 20  &  \refb  &   Keck    &   Yes     &  $0.076   \pm 0.003  $   &  $138.6 \pm 2  $   \\
2000 Jul 4   &  \refc  &   MMT     &   No      &  $0.078   \pm 0.003  $   &  $148   \pm 3  $   \\ 
2001 Mar 9   &  \refc  &   SAO     &   No      &  $0.0896  \pm 0.0010 $   &  $321   \pm 1  $   \\
2001 Mar 10  &  \refc  &   SAO     &   No      &  $0.0899  \pm 0.0010 $   &  $320   \pm 1  $   \\
2000 Aug 24  &  \refb  &   Keck    &   Yes     &  $0.059   \pm 0.001  $   &  $178.4 \pm 2  $   \\
2001 Jan 9   &  \refb  &   Keck    &   Yes     &  $0.073   \pm 0.002  $   &  $291.4 \pm 2  $   \\
2001 May 10  &  \refb  &   Keck    &   Yes     &  $0.097   \pm 0.001  $   &  $341.1 \pm 3  $   \\
2001 Jun 28  &  \refd  &   Keck    &   Yes     &  $0.1024  \pm 0.0012 $   &  $352.6 \pm 2  $   \\
2001 Sep  1  &  \refd  &   Keck    &   Yes     &  $0.1033  \pm 0.0010 $   &  $  9.7 \pm 2  $   \\
2002 Jun 26  &  \refe  &   HST     &   No      &  $0.090   \pm 0.003  $   &  $ 94.4 \pm 1.2$   \\
2003 Jul 11  &  \refe  &   Subaru  &   No      &  $0.085   \pm 0.007  $   &  $319.8 \pm 4.4$   \\
2004 Dec 24  &  \reff  &   Keck    &   Yes     &  $0.0885  \pm 0.0048 $   &  $111.3 \pm 1.2$   \\
2005 Feb 25  &  \reff  &   Keck    &   Yes     &  $0.0798  \pm 0.0029 $   &  $133.7 \pm 0.7$   \\
2008 Jan 16  &  \refg  &   Keck    &   No      &  $0.0618  \pm 0.0007 $   &  $272.5 \pm 1.4$   \\
2008 Jul 24  &  \refh  &   NOT     &   Yes     &  $0.0981  \pm 0.0022 $   &  $351.8 \pm 1.4$   \\
2009 Apr 29  &  \refg  &   Keck    &   No      &  $0.1002  \pm 0.0006 $   &  $ 66.5 \pm 0.6$   \\
2008 May 29  &  \refg  &   Keck    &   No      &  $0.1005  \pm 0.0018 $   &  $ 75.0 \pm 1.7$   \\
2009 Jun  4  &  \refh  &   WHT     &   Yes     &  $0.0984  \pm 0.0011 $   &  $ 78.2 \pm 0.6$   \\
2009 Jun 11  &  \refi  &   Keck    &   Yes     &  $0.099   \pm 0.002  $   &  $ 79.4 \pm 0.2$   \\ 
2010 Mar 22  &  \refg  &   Keck    &   No      &  $0.0558  \pm 0.0003 $   &  $206.5 \pm 1.1$   \\
2010 May 23  &  \refg  &   Keck    &   No      &  $0.0599  \pm 0.0004 $   &  $268.1 \pm 0.4$   \\ \hline
\end{tabular} 
 \begin{list}{}{}
 \item[References:] \refa\citet{MartinE:discv};\refb\citet{LaneB:orbbd}; \refc\citet{KenworthyM:Gl569Bay};\refd\citet{ZapateroM:dynmbb}; \refe\citet{ZapateroM:lithd};
   \reff\citet{SimonM:gl569ms}; \refg\citet{DupuyT:studpd}; \refh This work; \refi\citet{KonopackyQ:highpd}
 \item[Column ``In Fit''] flags the data points used in our orbital solution. Data points flagged with NO correspond to new additions appeared during the referee stage.
 \end{list}
\end{table}

\begin{table*}
\centering
\caption{Orbital Parameters of GJ569Bab.}
\label{tab:orbit}
\begin{tabular}{l c c c c c}
\hline \hline

Parameter                                  & Zapatero04$^{(1)}$   &    Simon06$^{(2)}$  & Konopacky10$^{(3)}$    & Dupuy10$^{(4)}$     & This Paper \\ \hline 

Semimajor axis, a (mas)                    & $92\pm 2$\refa      & $  90.4  \pm 0.7$  & $  90.8  \pm 0.8$      & $95.6^{+1.1}_{-1.0}$ & $ 90.1  \pm 0.7$   \\
Eccentricity, e                            & $0.32 \pm 0.01$     & $   0.312\pm 0.007$& $   0.310\pm 0.006$    & $0.316 \pm 0.005$  & $  0.317\pm 0.010$ \\
Inclination, i (\degr)                     & $ 34    \pm 2$      & $  32.4  \pm 1.3$  & $  33.6  \pm 1.3$      & $35.0  \pm 1.1$    & $ 30.0  \pm 1.6$   \\
Periapsis argument, $\omega$ (\degr)       & $257    \pm 2$      & $ 256.7  \pm 1.7$  & $  77.4  \pm 1.7$\refb & $257.9 \pm 2.0$    & $262.5  \pm 2.2$   \\
Ascending Node Longitude, $\Omega$ (\degr) & $321.5  \pm 2.0$    & $ 321.3  \pm 2.2$  & $ 144.8  \pm 1.9$\refb & $324.8 \pm 2.0$    & $313    \pm 4$     \\
Period, P(days)                            & $876    \pm 9$      & $ 863.7  \pm 4.2$  & $ 865.1  \pm 0.7 $     & $864.0 \pm 1.1$    & $870    \pm 9$     \\
Total Mass, M(\MS)\refc                    & $0.125\pm 0.005$    & $   0.125\pm 0.007$ & $   0.126\pm 0.007$   & $$                   & $0.122\pm 0.007$\\
Total Mass, M(\MS)\refd                    & $0.119\pm 0.005$    & $   0.119\pm 0.007$ & $   0.120\pm 0.007$   & $0.140^{+0.009}_{-0.008}$ & $0.116\pm 0.007$\\
Epoch, T(yrs)\refe                         & $2000.758\pm 0.008$ & $2000.754\pm 0.007$& $2003.150\pm 0.005$    & $2010.255\pm 0.004$&                    \\ 
Epoch, T(yrs)\reff                         & $2010.36 \pm 0.10 $ & $2010.22 \pm 0.05 $& $2010.260\pm 0.008$    & $2010.255\pm 0.004$& $2010.263\pm 0.012$\\ \hline
\end{tabular} 
\begin{list}{}{}
\item[References:] $^{(1)}$\citet{ZapateroM:dynmbb}, $^{(2)}$\citet{SimonM:gl569ms}, $^{(3)}$\citet{KonopackyQ:highpd}, $^{(4)}$\citet{DupuyT:studpd}
\item[Notes:]
\begin{list}{}{}
\item[]
\item[\refa] In original paper quoted in AU. Converted to angular distance assuming GJ569 Hipparcos parallax $\pi=0.10191 \pm 0.00167$ mas
  \citep{PerrymanM:HIPPARCOS}.

\item[\refb] $\omega$ and $\Omega$ derived solely from astrometry results exhibit a 180\degr degeneracy which is broken when using spectroscopic data
  (e.g. radial velocities) such as in \citet{ZapateroM:dynmbb}, \citet{SimonM:gl569ms} and \citet{KonopackyQ:highpd}. However in the latter work there
  seems to appear again the degeneracy which was broken in \citet{ZapateroM:dynmbb}. The reason why such a degeneracy remains in
  \citet{KonopackyQ:highpd} is unknown to us.

\item[\refc] Total Mass derived using the GJ569A Hipparcos parallax $\pi=0.10191 \pm 0.00167$ mas \citep{PerrymanM:HIPPARCOS} in original works by
  \citet{ZapateroM:dynmbb}, \citet{SimonM:gl569ms} and \citet{KonopackyQ:highpd}.

\item[\refd] Total Mass computed with the revised Hipparcos parallax $\pi=0.10359 \pm 0.00172$ mas \citep{vanLeeuwenF:hippnr} as pointed out in
  \citet{DupuyT:studpd}. The total mass estimates in \citet{ZapateroM:dynmbb}, \citet{SimonM:gl569ms} and \citet{KonopackyQ:highpd} scaled to the new
  parallax value.

\item[\refe] Epoch of periapsis appearing in original works.
\item[\reff] Epoch of periapsis propagated to fall within 2010 for the ease of comparison.
\end{list}
\end{list}
\end{table*}

In Table~\ref{tab:orbit}, in addition to the orbital parameters of the brown dwarf system, we also list the total mass estimates found in the
literature together with our estimation which is consistent but slightly lower than previous results.   

During the referee process of this work \citet{DupuyT:studpd} published a new set of values for the GJ569B system
astrometry. As the authors remark, there has been a recent revision by \citet{vanLeeuwenF:hippnr} of the GJ569 Hipparcos
parallax yielding a revised distance of $9.65 \pm 0.16$~pc instead of previous estimates of $9.81 \pm 0.16$~pc
\citep{PerrymanM:HIPPARCOS}; the impact of this new estimate of the GJ569 system distance is shown as an additional
entry in Table~\ref{tab:orbit}. \citet{DupuyT:studpd} claim previous astrometric values obtained with the Keck AO
facility in \citet{LaneB:orbbd,ZapateroM:dynmbb,SimonM:gl569ms} assumed a wrong plate scale. The same situation happens
for the HST value on 2002 June 26th \citep[which is also considered in][]{ZapateroM:lithd}. \citet{DupuyT:studpd}
reanalyze both the astrometric data in \citet{SimonM:gl569ms} and the HST value on 2002 June 26th, and together with new
5 data points from January 2008 until May 2010 a new orbit is computed providing a mass estimate for the GJ569B system
of $0.140^{+0.009}_{-0.008}$~\MS. The comparison of the new orbit by \citet{DupuyT:studpd} against previous orbits in
the literature is shown in a fraction of the orbit spanning the period April-June 2009 in
Fig.~\ref{fig:DataOrbitsZoom}. From theoretical models \citet{DupuyT:studpd} estimate the mass fraction parameter to be
$q\simeq 0.87 \pm 0.02$, which agree very well with observational results in \citet{ZapateroM:dynmbb} and
\citep{KonopackyQ:highpd}. The new astrometric values in \citet{DupuyT:studpd} have not been considered in our orbital
solution.

\subsection{Constraining the binary nature of GJ569Ba.}\label{SubSect:BinaryBa}

From radial velocities \citet{ZapateroM:dynmbb} assign individual masses of 0.071$\pm$0.011 $M_{\odot}$ and
0.054$\pm$0.011 $M_{\odot}$ for the GJ569Ba and GJ569Bb components, respectively. This is consistent with
0.073$\pm$0.008~\MS and 0.053$\pm$0.006~\MS by \citet{KonopackyQ:highpd} from radial velocity measurements. However,
upon the discovery of the multiple nature of GJ569B by \citet{MartinE:discv}, the possibility of GJ569Ba binarity was
discussed. The reason argued in \citet{MartinE:discv} was the failure of theoretical models to fit the observational
data with a single isochrone. The binary nature of GJ569Ba was again suggested in \citet{KenworthyM:Gl569Bay} on the
basis of Bb being about half as bright as Ba while the GJ569B spectrum was fitted by a single M8.5 component
and without evidence of reddening. \citet{SimonM:gl569ms} gives further support to the Ba binarity when accounting from
their estimated mass ratio $q=$\MBb/\MBa=0.19$\pm$0.13 as derived from their radial velocities.  \citet{SimonM:gl569ms}
suggest GJ569B is actually a triple system with Baa, Bab (i.e. the two suggested components of Ba) and Bb with nearly
the same masses of around 0.04-0.05~\MS. How does such a possibility fit into the larger collection of data since
\citet{SimonM:gl569ms}?

\begin{enumerate}

\item The flux ratio evolution with wavelength is not consistent with a system of nearly three-equal mass objects. In
  such a case the flux ratio should remain constant while its evolution from I-band to L-band is clearly decreasing:
  $0.622 \pm 0.017$, $0.51\pm0.02$, $0.590 \pm 0.023$, $0.504 \pm 0.015$ and $0.49 \pm 0.03$ in I, J, H, K and
  L$^\prime$ bands \citep[the L$^\prime$ measurement in][]{DupuyT:studpd}. This flux evolution is, however, consistent
  with Ba and Bb exhibiting slightly different radii (Ba larger than Bb) as suggested from theoretical models of BDs for
  the assumed masses of Ba and Bb and a slightly change in the effective temperature (Ba slightly hotter than Bb)
  consistent with the assigned spectral types to Ba and Bb in \citet{LaneB:orbbd}. Notice also that the absence of a Li
  feature in the combined spectrum of GJ569B in \citep{ZapateroM:lithd} plays against the three-equal mass objects
  hypothesis: assuming the highest mass estimate for the GJ569B system of $0.140^{+0.009}_{-0.008}$~\MS
  \citep{DupuyT:studpd} would inevitably ensure the presence of Li features as all objects would be below the 0.060~\MS
  threshold \citep{ChabrierG:evomv} but not such a feature was detected in \citet{ZapateroM:lithd}.

\item The age of the GJ569B is still an open issue with \citet{SimonM:gl569ms} assigning a similar age to the Pleiades
  ($\sim 120$~Myr) and the rest of estimates in the literature proposing ages in the range 200-800 Myr. The (I-J) colors
  of 2.72$\pm$0.08 and 2.83$\pm$0.08 for the Ba and Bb components, respectively, do not match with what is expected from
  Pleiades objects with masses $\sim 0.04-0.05$~\MS in the J vs. (I-J) Pleiades color-magnitude diagram in
  \cite{BihainG:Pleialm}. Note that the older the system the more inconsistent the measured (I-J) colors of Ba and Bb
  with respect to masses as low as 0.04-0.05~\MS. Note also that according to the the Lyon \citep{ChabrierG:evomv} and
  Tucson \citep{BurrowsA:Nongrayt} evolutionary models and for ages older than 250 Myr, objects with masses in the range
  0.04-0.05~\MS should exhibit spectral types in the L and T domain which is in contradiction with the M8.5-9.0 spectral
  types assigned to the unresolved GJ569B and resolved Ba and Bb components.

\end{enumerate}

\begin{figure}
  \centering
  \resizebox{\hsize}{!}{\includegraphics[angle=90,width=10cm]{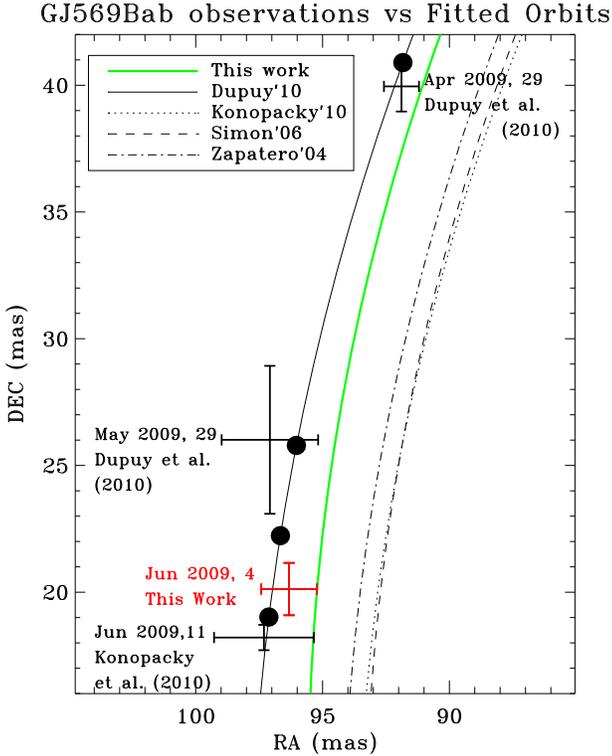}}
  \caption{Zoomed part of orbit showing orbit in period April-June 2009 together with orbits in the literature and
    predicted positions in orbit model by \citet{DupuyT:studpd}.}
  \label{fig:DataOrbitsZoom}
\end{figure}

However, there is still the possibility that Ba is composed of two components, namely Baa and Bab, of different
masses. For the purpose of the following discussion on the restrictions on the GJ569Ba binarity we assume the total mass
estimate of \MTot$=0.140^{+0.009}_{-0.008}$~\MS in \citet{DupuyT:studpd} and a mass ratio $q=0.80 \pm 0.15$ which lies
midway and consistently between the observed $q$ values in in \citet{ZapateroM:dynmbb} and \citet{KonopackyQ:highpd},
and the estimates by \citet{DupuyT:studpd} of $0.866^{+0.019}_{-0.014}$ and $0.886^{+0.021}_{-0.017}$ using the Lyon and
Tucson evolutinary models.  With these \MTot and $q$ values we have \MBa$=0.078 \pm 0.008$~\MS and and \MBb$=0.062 \pm
0.008$~\MS. A close look at the astrometric values reported in April-May 2009 by \citet{DupuyT:studpd}, by June 2009 in
\citet{KonopackyQ:highpd} and in this work are shown in Fig.~\ref{fig:DataOrbitsZoom}, allowing us to place a constrain
to the combination of mass and orbit semi-axis of a possible component Bab. From the above discussion we may safely
assume that if GJ569Ba is binary then its most massive component must satisfy 0.060 <\MBaa/\MS < 0.086 and $0 \le
$\MBab/\MS < 0.018. Also, if Bab exists it should be either very faint and orbit around Baa with a semi-axis smaller
than 0.5 AU (i.e. midpoint between Ba and Bb) and, using \MBa$=0.078 \pm 0.008$~\MS, a possible orbital period T<
1.3~yr. The presence of Bab would induce a motion in the position of GJ569Baa which would reflect itself as a wobble of
the orbit of GJ569Bb around the computed Keplerian orbits and with an amplitude $\alpha$ given by
\citep[e.g.][]{PerrymanM:exoplanets}:

\begin{equation}
\alpha = \frac{\mathrm{M}_{P}}{\mathrm{M}_{\star}} \, \frac{\mathrm{a}}{\mathrm{d}}
\end{equation}

\noindent where $\mathrm{M}_{P}$ and $\mathrm{M}_{\star}$ are the masses of GJ569Bab and GJ569Baa, respectively, a is
the Baa-Bab orbit semi-axis and d the distance to GJ569 (i.e. $d \sim 10$~pc). Examining in detail the deviation of
observed positions with respect to the \citet{DupuyT:studpd} orbit in the two month period depicted in
Fig.~\ref{fig:DataOrbitsZoom} we place an upper limit $\alpha$ < 2.2~mas with 1$\sigma$ confidence level.  Using \MBaa <
0.086~\MS we obtain the constrain $\mathrm{a}\cdot$\MBab <1.93, with \MBab in units of \MJ (Jupiter mass) and a in
AU. On the other hand we have placed an upper limit T< 1.3~yr on the period of Bab around of Baa. The residual rms of
the observed positions on the period April 2009--May 2010 with respect the predicted positions allows to place an upper
limit $\alpha$ < 1.9~mas with 1$\sigma$ confidence level and from here the constrain $\mathrm{a}\cdot$\MBab
<1.68. Adopting the former and more conservative constrain and since the minimum separation between GJ569Ba and GJ569Bb
is $\sim 0.5$ AU, the widest possible orbit of Bab around Baa would be < 0.25 AU and we would have \MBab
$\lesssim$~8~\MJ and the closer Bab to Baa the higher the feasible value for \MBab. Given the range of masses under
consideration and the distances between Baa and Bab, radial velocities studies of GJ569Ba should be able to shed further
light into the existence of the Bab component.

%

\section{Conclusions.}\label{Sect:Conclusions}

We have presented results showing the potential for high precision astrometry, differential photometry and high contrast
imaging using a Lucky Imaging instrument coupled to an adaptive-optics system. Our results indicate that with 4-m class
telescopes equipped with a moderately low order adaptive optics system it is possible to achieve angular resolutions
better than 0\farcs1 in the I-band. This is comparable to what is achieved in the Ks band with the use of AO-systems at
8-10 m class telescopes. This work is part of an effort to determine the feasibility of a Lucky Imaging instrument to be
coupled on the future AO system at the GTC telescope (GTCAO) and what the expectations in terms of high contrast imaging
and angular resolution should be expected from such a combination.

Our work has focused on the observation of the GJ569 system which contains a benchmark brown dwarf binary. The
application of our LM PSF fitting technique to the GJ569 image obtained with the WHT, where the GJ569B components are
resolved, allows to achieve high precision relative photometry (to a few millimagnitudes) and astrometry (to a few mas)
thanks to the availability of meaningful error bars associated to each pixel in the fitted image.

The potential for detection of faint companions has been addressed by looking at the $3\sigma$ detectability curves
in Fig.~\ref{fig:ContrastWHT}. On the images directly from the frame selection procedure, we distinguish two regions in
which the detectability behaves differently versus increasing the percentage of images being employed. In regions where
the image is dominated by the wide swallow halo of the primary PSF, the detectability is improved by restricting the
percentage of images being employed since in that way there is more energy in the core and less in the halo. At large
distances from the primary, the image is dominated by background/detector noise and the detectability is improved by
simply adding as many images as possible. On the wavelet-processed images, both in the inner and outer regions, we only
see a benefit on the increase of the percentage of images used in the frame selection. With the wavelet-processed images
we observe a magnitude gain in the inner region (1.7 magnitudes at 1\arcsec with respect to the non-processed image) but
far away from the primary no net gain as those parts of the image are dominated by background/detector noise.

We have measured a differential magnitude at I band between GJ569Ba and GJ569Bb $\Delta m_{BaBb}=0.622 \pm 0.017$. When
used in conjunction of $\Delta m_{BaBb}$ in the J, H and K bands by previous works, fits well with the spectral
determination of M8.5-M9 for the brown dwarf binary derived in the near-infrared by \citet{LaneB:orbbd}. Our results in
$I$-band and those in $J$, $H$ and $K$-bands in \citet{LaneB:orbbd} clearly indicate that Ba is brighter than Bb. This
together with the $I-J$ color favors a half subspectral class earlier for Ba than for Bb (see Fig.~\ref{fig:ColorIJ}).

The astrometric quality achieved with FastCam allows to locate two new points on the GJ569Bb orbit around GJ569Ba. The
orbits in the literature and the one derived including our points do not differ significantly and therefore the orbital
parameters are in perfect agreement with those previously published, although our mass estimate of 0.116$\pm$0.007~\MS
using the updated Hipparcos parallax distance of 9.65$\pm$0.16~pc in \citet{vanLeeuwenF:hippnr}. Our mass estimate of
the multiple GJ569B system is somewhat smaller, but within error bars of previously published values except for the
newly derived mass of $0.140^{+0.009}_{-0.008}$~\MS in \citet{DupuyT:studpd}. Our WHT data point on June 2009 falls
within 1 sigma from the \citet{DupuyT:studpd} orbital solutions and our own orbital solution although our NOT data point
on July 2008 is more consistent with previous orbital solutions in
\citet{ZapateroM:dynmbb,SimonM:gl569ms,KonopackyQ:highpd}. For a substantial refinement of the orbital parameters it
would be necessary to sample the whole orbit with similar uncertainties as those derived from our observation at the WHT
on June 2009.

The data available on the GJ569B system is consistent with a primary of $0.081 \pm 0.010$~\MS and a secondary of
$0.059\pm 0.007$~\MS. If Ba were a binary system then the Bab component would have a mass < 0.018 \MS. A simple
qualitative analysis on the deviations from a Keplerian orbit allows to place an upper limit to the product of mass and
orbit semi-axis of this object.


\section*{Acknowledgments}
 B.F. and L.L. are funded by the Spanish MICINN under the Consolider-Ingenio 2010 Program gran CSD2006-00070: First
 Science with the GTC (\url{http://www.iac.es/consolider-ingenio-gtc}).  The 0.82-m IAC80 Telescope is operated on the
 island of Tenerife by the Instituto de Astrofisica de Canarias in the Spanish Observatorio del Teide. This research has
 made use of the Washington Double Star Catalog maintained at the U.S. Naval Observatory. We thank the referee
 Mar\'{\i}a Rosa Zapatero Osorio for many insightful discussions which have substantially improved the contents of the
 orginal manuscript. We also wish to thank to the IAC teams lead by Emilio Cadavid and Vicente S\'anchez and all the ING
 staff, and specially to Chris Benn and Tibor Agocs, for their very precious help during the setup and observations with
 FastCam at the WHT.

%

\bibliographystyle{mn2e}
\bibliography{/home/bfemenia/bibliografia/bibtex/ao,/home/bfemenia/bibliografia/bibtex/AOLI_Science,/home/bfemenia/bibliografia/bibtex/BD_VLM,/home/bfemenia/bibliografia/bibtex/others}

\label{lastpage}

\end{document}